\documentstyle[preprint,aps]{revtex}
\begin{document}
\preprint{UMHEP-427}
\draft
\title{Systematics of soft final state interactions 
in $B$ decay}
\author{John F. Donoghue, Eugene Golowich,
Alexey A. Petrov and Jo\~ao M. Soares} 
\address{Department of Physics and Astronomy, 
University of Massachusetts, Amherst MA 01003 USA}
\maketitle
\begin{abstract}
\noindent 
By using very general and well established features of 
soft strong interactions we show, contrary to 
conventional expectations, that (i) soft final state interactions 
(FSI) do not disappear for large $m_B$, (ii) inelastic rescattering 
is expected to be the main source of soft FSI phases, and (iii) flavor 
off-diagonal FSI are suppressed by a power of $m_B$, but are quite 
likely to be significant at $m_B\simeq 5$~GeV.  We briefly discuss the 
influence of these interactions on tests of CP-violation and on 
theoretical calculations of weak decays.  
\end{abstract}
 
It is notoriously difficult to say anything useful about final state
interactions in weak decays.  Although the final state interactions are
not themselves of fundamental interest, they are
important for some truly interesting aspects of B decay.  For
example, many signals of direct CP violation in B transitions require 
final state phases as well as CP-violating phases if the CP-odd
asymmetry is to be nonzero.$^{\cite{dgh}}$  In this paper we shall derive 
some general properties of soft final state interactions 
and describe the implications for theory and phenomenology. 

The scattering of hadrons at high energies exhibits a two-component 
structure of `soft' and `hard' scattering.  Soft scattering is that 
which occurs primarily in the forward direction.  The transverse
momentum is limited, having a distribution which falls exponentially 
with a scale of order $0.5$~GeV.  At higher transverse momentum
ultimately, one encounters the region of hard scattering, which 
falls only as a power of the transverse momentum.  Collisions
involving hard scattering are interpreted as interactions between 
pointlike constituents of the hadrons, the quarks and gluons 
of QCD. These are calculable in QCD perturbation theory and are found 
to be in good quantitative agreement with experiment.  Hard scattering
is, however, only a very small portion of the total hadronic cross
section.  The much larger soft component at low values of transverse 
momentum is by far the dominant contribution to high energy
scattering.  Although soft hadronic interactions are generally not
calculable from first principles, there is available a wealth 
of experimental studies$^{\cite{pdg}}$ and accurate high energy 
phenomenology$^{\cite{pl}}$ on which to base our study.  

The modern approach to $B$ physics employs as an organizing
principle the fact that the $B$ mass is very large compared to the QCD scale.  
In the context of soft FSI in $B$ decays, it suggests the question ---
what is the leading order behavior of soft final state phases 
in the $m_B \to \infty$ limit?  
The common perception among researchers is that they should become 
less and less important as the mass of the decaying quark becomes
heavier.  This is because, roughly speaking, `the final state particles
emerge at such high momenta that they do not have a chance to
rescatter'.  Such an expectation is, however, false because soft
scattering actually grows with energy.  As an example of this
important energy dependence, we shall demonstrate below
that the imaginary part of the forward elastic amplitude 
has an $s^{1 + \eta}$ ($\eta \simeq 0.08$) dependence, and as a
consequence, the elastic final state interaction is roughly constant
as a function of $m_B$.  We shall then use this observation as the starting 
point for a more general exploration of the systematics of FSI for 
large $m_B$.  The inevitability of our conclusions will be seen to 
follow rather directly from well established aspects of 
strong interaction phenomenology.  

Final state interactions in B decay involve the rescattering of
physical final state particles.  Unitarity of the
${\cal S}$-matrix, ${\cal S}^\dagger {\cal S} = 1$, implies 
that the ${\cal T}$-matrix, ${\cal S} = 1 + i {\cal T}$, obeys 
\begin{equation}
{\cal D}isc~{\cal T}_{B \rightarrow f} \equiv {1 \over 2i} 
\left[ \langle f | {\cal T} | B \rangle - 
\langle f | {\cal T}^\dagger | B \rangle \right] 
= {1 \over 2} \sum_{I} \langle f | {\cal T}^\dagger | I \rangle 
\langle I | {\cal T} | B \rangle \ \ .
\label{unit}
\end{equation}
Of interest are all physical intermediate states which can scatter
into the final state $f$.  Among all these, however, we shall first 
concentrate on just the {\it elastic} channel and 
demonstrate that elastic rescattering does not disappear 
in the limit of large $m_B$.\footnote{We stress that we are {\it
not} suggesting the elastic channel to be the dominant contribution 
to soft rescattering.  Our analysis leads to quite the opposite 
conclusion, that it is the inelastic channels which are most important.}
The elastic channel is especially
convenient for our discussion because we can use the optical theorem 
to rigorously connect it to known physics.  The optical theorem
relates the forward invariant amplitude ${\cal M}$ to the total 
cross section, 
\begin{equation}
{\cal I}m~{\cal M}_{f\to f} (s, ~t = 0) = 2 k 
\sqrt{s} \sigma_{f \to {\rm all}} \sim s \sigma_{f \to {\rm all}} \ \ ,
\label{opt}
\end{equation}
where $s$ is the squared center-of-mass energy and $t$ is the squared
momentum transfer. 

The asymptotic total cross sections are known
experimentally to rise slowly with energy.  All known cross sections
can be parameterized by fits of the form$^{\cite{ld}}$ 
\begin{equation}
\sigma (s) = X \left({s\over s_0}\right)^{0.08} 
+ Y \left({s\over s_0}\right)^{-0.56} \ \ ,
\label{pl}
\end{equation}
where $s_0 = {\cal O}(1)$~GeV is a typical hadronic scale.  
Thus, the imaginary part of the forward elastic scattering amplitude 
rises asymptotically as $s^{1.08}$.  This growth with s is
counterintuitive in that it cannot be generated by a perturbative
mechanism at any finite order.  In particular, calculations based on
the quark model or perturbative $QCD$ would completely miss this
feature.

In order to arrive most simply at our
goal, let us first consider only this imaginary part, and build in 
the known exponential fall-off of the elastic cross section in $t$ 
(recalling that $t$ is negative) by writing 
\begin{equation}
i{\cal I}m~{\cal M}_{f\to f} (s,t) \simeq i \beta_0 \left( {s \over s_0}
\right)^{1.08} e^{bt} \ \ .
\label{fall}
\end{equation}
It is then an easy task to calculate the contribution 
of the imaginary part of the elastic amplitude to the
unitarity relation for a final state $f = a + b$ with kinematics 
$p_a' +  p_b' = p_a +  p_b$ and $s = (p_a + p_b )^2$, and we find 
\begin{eqnarray}
{\cal D}isc~{\cal M}_{B \to f} &=&
{1 \over 2} \int {d^3p_a' \over (2\pi)^3 2E_a'}
{d^3p_b' \over (2\pi)^3 2E_b'}
(2 \pi)^4 \delta^{(4)} (p_B - p_a' - p_b') \cdot -i\beta_0 
\left( {s \over s_0} \right)^{1.08} e^{b(p_a - p_a')^2} 
{\cal M}_{B \rightarrow f} \nonumber \\
&=& - {1\over 16\pi} {i\beta_0 \over s_0 b}\left( {m_B^2 \over s_0} 
\right)^{0.08} {\cal M}_{B \rightarrow f} \ \ ,
\label{mess}
\end{eqnarray}
where we have used $t = (p_a - p_a')^2 \simeq 
-s(1 - \cos\theta)/2$ and have taken $s = m_B^2$. 
The integration over the angle involving the direction of the 
intermediate state is seen to introduce a suppression factor 
to the final state interaction of $s^{-1} = m_B^{-2}$.  
This is because the soft final state rescattering can take place only 
if the intermediate state has a transverse momentum $p_\perp 
\le 1$~GeV with respect to the final particle direction.  This would 
naively suggest a result consistent with conventional 
expectations, {\it i.e.} an FSI which falls as $m_B^{-2}$.  However,
the fact that the forward scattering amplitude {\it grows} with a
power of $s$ overcomes this suppression and leads to elastic 
rescattering which does not disappear at large $m_B$.  

In fact, we can make a more detailed estimate of elastic 
rescattering because the phenomenology of high energy 
scattering is well accounted for by Regge theory.$^{\cite{jc}}$  
Scattering amplitudes are described by the exchanges of 
Regge trajectories (families of particles of differing spin) 
which lead to elastic amplitudes of the form 
\begin{equation}
{\cal M}_{f \rightarrow f} = \xi \beta (t) 
\left( {s \over s_0} \right)^{\alpha (t)} e^{i \pi \alpha(t)/2}
\label{regge}
\end{equation}
with $\xi = 1$ for charge conjugation $C=+1$ and 
$\xi = i$ for $C=-1$. Each such trajectory is described by a straight line, 
\begin{equation}
\alpha (t) = \alpha_0 + \alpha' t \ \ .
\label{traj}
\end{equation}
The leading trajectory for high energy scattering is the 
Pomeron, having  $C=+1, \alpha_0 \simeq 1.08$ and 
$\alpha' \simeq 0.25$~GeV$^{-2}$.  Note that since 
\begin{equation}
\left( {s \over s_0} \right)^{\alpha (t)} = 
\left( {s \over s_0} \right)^{\alpha_0} e^{\alpha' 
\ln\left(s /s_0\right)~t}\ \ ,
\label{slope}
\end{equation}
the exponential fall-off in $t$ is connected with the slope $\alpha'$
and the effective slope parameter $b$ in Eq.~(\ref{fall}) 
thus increases logarithmically with $s$.  Since $\alpha_0$ is near
unity, the phase of the Pomeron-exchange amplitude is seen from
Eq.~(\ref{regge}) to be almost purely imaginary.  This feature 
has been verified experimentally by interference measurements.  There 
are several next-to-leading trajectories, both those with $C=-1$ 
($\rho(770)$ $\&$ $\omega(782)$ trajectories) and those with $C=+1$ 
($a_2 (1320)$ $\&$ $f_2(1270)$ trajectories).  Roughly, these have 
$\alpha_0 \simeq 0.44$, $\alpha' \simeq 0.94$~GeV$^{-2}$ and lead 
collectively to the $s^{-0.56}$ 
dependence in the asymptotic cross section of Eq.~(\ref{pl}). 
The prefactor $\beta(t)$ in Eq.~(\ref{regge}) also has known 
regularities.  For the Pomeron, $\beta$ is very nearly proportional
to the number of quarks at each vertex, and carries a power law 
behavior similar to the electromagnetic form factor.  
Therefore, $\beta_{\pi\pi}$ in pion-pion scattering can be expressed
in terms of the analogous proton-proton quantity $\beta_{pp}$ as 
\begin{equation}
\beta_{\pi\pi} (t) = \left( {2\over 3} \right)^2 
{\beta_{pp} (t = 0) \over (1 - t /m_\rho^2)^2} \ \ .
\label{res}
\end{equation}
The combination of exponential and power law $t$
dependence in a generic Regge amplitude gives a unitarity integral 
no longer having an elementary form.  However, the integration 
can still be carried out in terms of Euler functions.  Taking 
$s = m_B^2 \simeq 25~{\rm GeV}^2$, we obtain for the Pomeron
contribution 
\begin{equation}
{\cal D}isc~{\cal M}_{B \to \pi\pi}|_{\rm Pomeron} = -i\epsilon
{\cal M}_{B \to \pi\pi} \ \ ,
\label{despite}
\end{equation}
where we find from our computation, 
\begin{equation}
\epsilon \simeq 0.21 \ \ .
\label{eps}
\end{equation}
>From this numerical result and from the nature of its derivation, 
we may anticipate that additional individual soft FSI will not be vanishingly 
small.  Moreover, other final states should have elastic rescattering effects 
of comparable size.  However, of chief significance is the weak 
dependence of $\epsilon$ on $m_B$ that we have found --- the
$(m_B^2)^{0.08}$ factor in the numerator is attenuated by the 
$\ln(m_B^2/s_0)$ dependence in the effective value of $b$ (compare 
Eqs.~(\ref{fall}),(\ref{slope})). 

The above study of the elastic channel, although instructive, is far 
from the whole story.  In fact, it suggests the 
even more significant result that at high energies {\it FSI phases 
are generated chiefly by inelastic effects}. At a physical level, 
this conclusion is forced on us by the fact that
the high energy cross section is mostly inelastic. It is also 
plausible at the analytic level, given that the Pomeron elastic amplitude is 
almost purely imaginary.  The point is simply this.  Our study of 
elastic rescattering has yielded a ${\cal T}$-matrix element ${\cal T}_{ab
\to ab} = 2 i \epsilon$, which directly gives ${\cal S}_{ab \to ab} = 1- 2 
\epsilon$. However, the constraint of the ${\cal S}$-matrix be unitary 
can be shown to imply that the 
off-diagonal elements must be ${\cal O}(\sqrt{\epsilon})$. 
Since $\epsilon$ is 
approximately ${\cal O}(m_B^0)$ in powers of $m_B$ and numerically 
$\epsilon < 1$, the inelastic amplitude must also be ${\cal
O}(m_B^0)$ and of magnitude $\sqrt{\epsilon} > \epsilon$.  
There is an alternate argument, 
utilizing 
the form of the final state unitarity relations, which also 
shows that inelastic effects are required to be present.  
In the limit of T-invariance for the weak interactions, the discontinuity 
${\cal D}isc~ {\cal{M}}_{B \to f}$ is a real number (up to irrelevant rephasing
invariance of the $B$-state). The factor of $i$ obtained in the elastic
rescattering in Eq.~(\ref{despite}) must be compensated for by the
inelastic rescattering (this effect is made explicit in the example 
to follow) in order to make the total real.  Therefore, the
presence of inelastic effects is seen to be necessary.  

Analysis of the final-state unitarity relations in their most general form, 
\begin{equation}
{\cal D}isc~{\cal{M}}_{B \to f_1} = {1\over 2}\sum_k\ {\cal M}_{B\to
k} T^\dagger_{k \to f_1} \ \ ,
\label{unit2}
\end{equation}
is quite complicated due to the many contributing intermediate states 
present at the $B$ mass.  However, it is possible to illustrate 
the systematics of inelastic scattering by means of a simple 
two-channel model.  This pedagogic example involves a two-body final state 
$f_1$ undergoing elastic scattering and a final state $f_2$ which is meant to
represent `everything else'.  We assume that the elastic amplitude 
is purely imaginary.  Thus, the scattering can be described 
in the one-parameter form 
\begin{equation}
 S = \pmatrix{\cos 2 \theta & i \sin 2 \theta \cr
              i \sin 2 \theta & \cos 2 \theta \cr} \ ,\qquad \qquad 
 T = \pmatrix{2 i \sin^2 \theta &  \sin 2 \theta \cr
               \sin 2 \theta & 2 i \sin^2 \theta \cr} \ \ ,
\label{matr1}
\end{equation}
where, from our elastic-rescattering calculation, we identify $\sin^2 
\theta \equiv \epsilon$. The unitarity relations become
\begin{eqnarray}
{\cal D}isc ~{\cal{M}}_{B \to f_1} = - i \sin^2 \theta {\cal{M}}_{B \to f_1} +
\frac{1}{2} \sin 2 \theta {\cal{M}}_{B \to f_2} \ \ ,\nonumber \\
{\cal D}isc~ {\cal {M}}_{B \to f_2} = \frac{1}{2} \sin 2 \theta 
{\cal{M}}_{B \to f_1} - i \sin^2 \theta {\cal{M}}_{B \to f_2} \ \ 
\label{big}
\end{eqnarray}
If, in the limit $\theta \to 0$, the decay amplitudes become the real numbers 
${\cal{M}}_1^0$ and ${\cal{M}}_2^0$, these equations are solved by
\begin{equation}
{\cal{M}}_{B \to f_1} = \cos \theta {\cal{M}}_1^0 + i \sin \theta
{\cal{M}}_2^0  \ , \qquad 
{\cal{M}}_{B \to f_2} = \cos \theta {\cal{M}}_2^0 + i \sin \theta
{\cal{M}}_1^0 \ \ .
\label{soln}
\end{equation}
As a check, we can insert these solutions back into Eq.~(\ref{big}). 
Upon doing so and bracketing contributions from ${\cal{M}}_{B \to
f_1}$ and ${\cal{M}}_{B \to f_2}$ separately, we find 
\begin{equation}
{\cal D}isc~{\cal{M}}_{B \to f_1} = {1\over 2}\left[ \bigg( -2i\epsilon 
{\cal M}^0_{B\to f_1} 
+ {\cal O}(\epsilon^{3/2}) \bigg) + \bigg( 2\sqrt{\epsilon} 
{\cal M}^0_{B\to f_2} + 2i\epsilon {\cal M}^0_{B\to f_1} \bigg) \right]\ \ .
\label{check}
\end{equation}
The first of the four terms comes 
from the elastic channel $f_1$ and is seen to be 
cancelled by the final term, which arises from the inelastic channel
$f_2$.  The third term is dominant, being  ${\cal O}(\sqrt{\epsilon})$, 
and comes from the inelastic channel.  

In this example, we have seen that the phase is given by the inelastic
scattering with a result of order 
\begin{equation}
\frac{ {\cal I}m~ {\cal{M}}_{B \to f}}{{\cal R}e~ {\cal{M}}_{B \to f}} \sim 
\sqrt{ \epsilon}~ \frac{{\cal{M}}_2^0}{{\cal{M}}_1^0} \ \ .
\end{equation}
Clearly, for physical $B$ decay, we no longer 
have a simple one-parameter ${\cal S}$ matrix.  However, the main
feature of the above result is expected to remain --- that 
inelastic channels cannot vanish because they 
are required to make the discontinuity real and that 
the phase is systematically of order $\sqrt{\epsilon}$ from these
channels.  Of course, with many channels, cancellations or enhancements are 
possible for the sum of many contributions. However the generic expectation
remains --- that inelastic soft final-state-rescattering arising
from Pomeron exchange will generate a phase which does not vanish in
the large $m_B$ limit.

What about nonleading effects? It is not hard to see that 
these may be significant at the physical values of $m_B$. For example, 
the fit to the $\bar p p$ total cross section is
\begin{equation}
\sigma ( p \bar p ) = \Bigl [ 22.7 \Bigl ( \frac{s}{s_0} \Bigr )^{0.08} +
140 \Bigl ( \frac{s}{s_0} \Bigr )^{-0.56} \Bigr ] ~~ (mb)
\end{equation}
with $s_0 = 1~{\rm GeV}^2$. At $s=(5.2~{\rm GeV})^2$, the nonleading 
coefficient is a factor of six larger that the leading effect,
effectively compensating for the $s^{-0.56}=m_B^{-1.12}$ suppression.
The subleading terms are then comparable in  
the elastic forward $\bar p p $ scattering amplitude. The slope of the 
$\rho$ trajectory and hence the experimental fall-off with $t$, is
larger than that of the Pomeron by a factor of nearly four, and thus 
this moderates 
the integrated rescattering effects. If we estimate the $\beta$ coefficient of
the $\rho$ trajectory in $\pi \pi$ by relating it to $\bar p p$ via a factor 
of $\beta_{\pi \pi} \simeq 4 \beta_{\bar p p} $ and then perform the 
integration over the intermediate state momentum we find
\begin{equation}
Disc~{\cal{M}}_{B \to \pi\pi} \bigg|_{\rho - {\rm traj}} = 
i \epsilon_\rho {\cal{M}}_{B \to \pi\pi} \ \ ,
\end{equation}
with $\epsilon_\rho \simeq 0.11 - 0.05~i$.  It is likely that the 
$f_2 (1270)$ trajectory could be somewhat larger, as it is in 
$\bar p p$ and $\pi p$ scattering.  

Final state phases can contribute to weak decay phenomenology 
in a variety of ways.  Here, we briefly consider two of 
these, isospin sum rules and CP-violating asymmetries.  A 
simple example of an isospin sum rule is the following relation 
between $B \to \pi \pi$ decay amplitudes, 
\begin{equation}
{\cal{M}}_{+-} - {\cal{M}}_{00} = {2\sqrt{2}\over 3}{\cal{M}}_{-0} \ \ ,
\end{equation}
where ${\cal{M}}_{+-} \equiv {\cal{M}} (B^0 \to \pi^+ \pi^-)$, {\it
etc}.  Measurement of the magnitude of each amplitude via the partial 
decay rate allows one to test the sum rule.  Noting that the $\pi\pi$ 
final state in $B$ decay occurs in the isospin states $I = 0,2$, one 
can solve for the difference in phase angles, 
\begin{equation}
\cos(\delta_0 - \delta_2) = {9 \over 4} \cdot
{|{\cal M}_{+-}|^2 - |{\cal M}_{00}|^2 \over 
|{\cal M}_{-0}| \sqrt{ 9|{\cal M}_{+-}|^2 + 
9|{\cal M}_{00}|^2 - 4|{\cal M}_{-0}|^2 }} 
\end{equation}
At a theoretical level, one sees that the leading Pomeron effect does
not contribute to these isospin sum rules since Pomeron exchange is 
identical for each  $\pi^i \pi^j$ final state and thus generates 
only a common overall phase.  Thus, the phases measured in isospin sum
rules are technically subleading, of order $m_B^{-1.12}$.

CP-violating asymmetries involve comparisions of $B \to f$ and
$\bar B \to \bar f$. In order to be nonzero, these require two 
different pathways to reach the final state $f$, and these two paths must
involve different CP-violating weak phases and different strong phases.
The leading Pomeron phases {\it can} contribute to such asymmetries if 
the other conditions are met. Because the strong phase is generated by 
inelastic channels, the relevant pathways would involve $B \to f$ directly
or $B \to$ `multibody' followed by the inelastic rescattering, `multibody' $\to f$.
Depending on the dynamics of weak decay matrix elements, these may pick
up different weak phases. As an example, consider the final state 
$f = K^- \pi^0$, which can be generated either by a standard $W$ exchange or
by the penguin diagram, involving different weak phases.$^{\cite{lw}}$   For the strong
rescattering, we must also consider a channel to which $K^- \pi^0$ 
scatter inelastically, which we call $K n \pi$ (although one can generate this 
asymmetry by a hard rescattering $D_s D \to K^- \pi^0$, we
are concentrating here on the soft physics). The $W$-exchange and penguin 
amplitudes will contribute with different weight to $K \pi$ and  
$K n \pi$, so that in the absence of final state interactions we expect
\begin{eqnarray}
{\cal{M}} (B^- \to K^- \pi) = |A_1| e^{i \phi_1} =
A_1^w e^{i \phi_w} + A_1^p e^{i \phi_p} \nonumber \\
{\cal{M}} (B^- \to K^- n \pi) = |A_n| e^{i \phi_n} =
A_n^w e^{i \phi_w} + A_n^p e^{i \phi_p}
\end{eqnarray}
with $\phi_1 \not = \phi_n$. If we now model the strong rescattering by the 
two channel model described above, we have for $B$ and $\bar B$ decays
\begin{eqnarray}
{\cal{M}} (B^- \to K^- \pi) = |A_1| e^{i \phi_1} +
i \sqrt{ \epsilon}~ |A_n| e^{i \phi_n} \nonumber \\
{\cal{M}} (B^+ \to K^+ n \pi) = |A_1| e^{-i \phi_1} +
i \sqrt{ \epsilon} ~|A_n| e^{-i \phi_n} 
\end{eqnarray}
This leads to a CP-violating decay rate asymmetry
\begin{equation}
\Gamma (B^- \to K^- \pi^0) - \Gamma (B^+ \to K^+ \pi^0) \sim
\sqrt{\epsilon} |A_1||A_n| \sin( \phi_n - \phi_1)
\end{equation}
While this effect will be very difficult to calculate, we see that inelastic 
final state interactions can contribute to CP-violating asymmetries at
leading order in $m_B$.

The results obtained in this paper must also be accounted for in any
theoretical calculation of weak decay amplitudes. For large $m_B$, 
there is the hope that one can directly calculate the weak matrix
elements through variants of the factorization hypothesis or by 
perturbative QCD.  Final state interactions will impose limits on 
the accuracy of such methods, as no existing technique includes the 
effect of inelastic scattering.  There must exist, in every valid 
theoretical calculation, a region of the parameter space where 
the nonperturbative Regge physics is manifest.  Arguments based on
local quark-hadron duality do not account for these effects of 
soft physics because the growth of the scattering amplitude with $s$ 
(for both the leading and first nonleading trajectories) cannot be 
seen in perturbative calculations.  It remains an intriguing possibility  
that the assumption of quark-hadron duality can be questioned in 
other aspects of $B$-decay as well.  At any rate, for final state 
interaction studies, one may only hope that the
perturbative/calculable physics is larger then the difficult
nonperturbative contributions discussed in this paper.  

To conclude, we have argued that the general features of soft
scattering have forced upon us some suprising conclusions regarding 
final state interactions.  Most importantly, the growth of forward
scattering with $s$, as required by the optical theorem and cross
section data, indicates that soft scattering does not decrease for 
large $m_B$. The structure of the elastic rescattering via the Pomeron
also requires that {\it inelastic} processes are the leading sources 
of strong phases. These systematics can be important for the
phenomenology of $B$ decays.

\vspace{0.5cm}

\vfill \eject 

\begin{references}

\bibitem{dgh} For example, see Section~5 of Chapter~XIV in 
J.F. Donoghue, E. Golowich and B.R. Holstein, {\it Dynamics of 
the Standard Model}, (Cambridge University Press, Cambridge, England 1992).
 
\bibitem{pdg} Particle Data Group, L.\ Montanet {\it et al.}, 
Phys. Rev. {\bf D50} (1994) 1173.  

\bibitem{pl} P.V. Landshoff, in {\it QCD - 20 Years Later}, 
(World Scientific, Singapore 1993).

\bibitem{ld} A. Donnachie and P.V. Landshoff, Phys. Lett. 
{\bf B296} (1992) 227.

\bibitem{jc} For example, see P.D.B. Collins, {\it Introduction to 
Regge Theory and High Energy Physics}, (Cambridge University Press, 
Cambridge, England 1977).
 
\bibitem{lw} Lincoln Wolfenstein, Phys. Rev. {\bf D43} (1991) 151. 


\end{references}
\end{document}